%% file: iwoce2009.tex
\documentclass{sig-alternate}
\usepackage{amssymb}
\usepackage{graphicx}
\usepackage{fp}
\usepackage{color}
\usepackage{url}
\usepackage{framed}
\usepackage{lastpage}
\usepackage{listings}
\usepackage{alltt}
\usepackage{amssymb}
\usepackage{amstext}
\usepackage{amsmath}
\usepackage{AMMALanguages}
\usepackage{xspace}
\usepackage{changebar}
\usepackage{hyperref}

\input{macros}

\begin{document}

\title{Towards Maintainer Script Modernization\\ in FOSS
  Distributions\titlenote{Partially supported by they European
    Community's 7th Framework Programme (FP7/2007--2013),
    \href{http://www.mancoosi.org}{\MANCOOSI} project, grant agreement
    n. 214898.}}

\numberofauthors{2}
\author{
\alignauthor
Davide {Di Ruscio}, Patrizio Pelliccione, and Alfonso Pierantonio\\
  \affaddr{Universit\`a degli Studi dell'Aquila}\\
  \affaddr{Dipartimento di Informatica}\\
  \email{\{davide.diruscio, patrizio.pelliccione, alfonso.pierantonio\}@univaq.it}
\alignauthor
Stefano Zacchiroli\\
  \affaddr{Universit\'e Paris Diderot}\\
  \affaddr{PPS, UMR 7126, France}\\
  \email{zack@pps.jussieu.fr}
}

\maketitle
\begin{abstract}
Free and Open Source Software (\FOSS) distributions are complex
software systems, made of thousands \emph{packages} that evolve
rapidly, independently, and without centralized coordination. During
packages \emph{upgrades}, corner case failures can be encountered and
are hard to deal with, especially when they are due to misbehaving
\emph{maintainer scripts}: executable code snippets used to finalize
package configuration.

In this paper we report a software modernization experience, the
process of representing existing legacy systems in terms of models,
applied to \FOSS{} distributions. We present a process to define
meta-models that enable dealing with upgrade failures and help rolling
back from them, taking into account maintainer scripts. The process
has been applied to widely used \FOSS{} distributions and we report
about such experiences.
\end{abstract}


\input{introduction}

\input{foss}
\input{upgrade}
\input{analysis}

\input{metamodel}
\input{related}
\input{conclusions}

\bibliographystyle{plain}
\bibliography{mancoosi}

\end{document}

%% file: macros.tex
\long\def\ignore#1{\relax}

\newcommand{\MANCOOSI}{\textsc{Mancoosi}}
\newcommand{\FOSS}{FOSS}
\newcommand{\SPEC}{\texttt{spec}}
\newcommand{\CIRCREF}[1]{\text{\textcircled{\raisebox{-0.8pt}{#1}}}}

\newcommand{\MAINTSCRIPT}[1]{\texttt{#1}}
\newcommand{\PRERM}{\MAINTSCRIPT{prerm}}
\newcommand{\PREINST}{\MAINTSCRIPT{preinst}}
\newcommand{\POSTINST}{\MAINTSCRIPT{postinst}}
\newcommand{\POSTRM}{\MAINTSCRIPT{postrm}}
\newcommand{\CONFIG}{\MAINTSCRIPT{config}}

%% file: introduction.tex
\begin{figure*}[t]
  \center
  \includegraphics[width=0.90\textwidth]{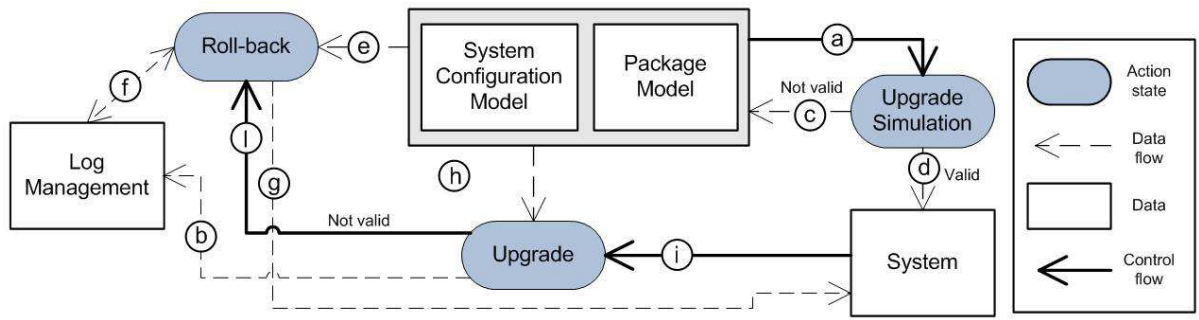}
  \caption{Model-driven approach to manage system configuration
    proposed}
  \label{fig:overallApproach}
\end{figure*}

\section{Introduction}\label{sec:introduction}

Free and Open Source Software (\FOSS) distributions, as well as other
complex systems, provide their software components in ``packaged''
form. Packages, available from remote repositories, are installed and
removed on local machines by means of \emph{package manager}
applications, such as APT~\cite{apt-howto} or Apache
Maven~\cite{apache-maven}. Package managers are responsible of both
finding suitable upgrade strategies by solving dependencies and
conflicts among packages, and of actually deploying the involved
packages on the filesystem, possibly aborting the operation if
problems are encountered.

During the installation and removal of a package, additional actions
are required in order to finalize the component within the overall
system configuration. Such actions are usually delegated to
executable \emph{maintainer scripts}, contained in the packages.
Maintainer scripts are written in fully general POSIX shell script
that makes very hard, impossible in the general case, to predict a
priori their side-effects which can affect the entire system. As a
consequence, a satisfactory solution able to deal with automatic
recovery of faults caused by misbehaving maintainer scripts is still
missing~\cite{hotswup:package-upgrades}.

Model Driven Engineering (MDE)~\cite{Bezivin05} can be crucial to
improve the management of system configurations since models can make
explicit dependencies and effects that are naturally
implicit. Representing a \FOSS{} installation with models paves the
way to two different kinds of support to upgrade management:
\begin{enumerate}
\item \emph{dry run} simulation of upgrades, looking for inconsistent
  configurations induced by misbehaving maintainer scripts or
  otherwise buggy packages;
\item fine-grained \emph{logging} of actions executed on the real
  installation during package deployment; the obtained log can then be
  used to better drive downstream rollback mechanisms.
\end{enumerate}


\noindent
A model is obviously an abstraction of the reality. In modeling it
is of crucial importance the level of abstraction taken into
account. On one side we have to abstract away many details in order
to have tractable models, on the other side the models must be able
to effectively present relevant maintainer scripts.
This paper faces with the precise problem of \emph{modernize} maintainer scripts focusing on
\emph{software modernization} of \FOSS{} installations and in particular of maintainer scripts. The
idea of renewing legacy systems by means of model driven approaches has been pursued by the Object
Management Group (OMG) since 2003. In particular, OMG defined the Architecture-Driven Modernization
(ADM) task force~\cite{ADM} to support software modernization of existing assets which are imported
into MDE enabled development environments.


In this paper, we analyze the domain of package-based \FOSS{}
distributions and formalize as meta-models installations, for the
purpose of upgrade simulation and logging. In particular we highlight
the analysis of maintainer scripts that has been conducted for the
Debian GNU/Linux and for some RPM-based distributions. The resulting
metamodels underpin the extraction process (also called
\emph{injection}) of models from existing \FOSS{} distributions
enabling the application of model-driven techniques and tools.

\paragraph{Paper structure}
We begin by providing necessary details about \FOSS{} distributions
in Section~\ref{sec:foss}. Section~\ref{sec:upgrade} outlines the
model-driven approach to deal with upgrade simulation and logging.
Section~\ref{sec:analysis} describes the analysis of maintainer
scripts on real-life distributions, while
Section~\ref{sec:elementsToBeModeled} describes the specification of
sample real maintainer scripts by using the defined modeling
constructs. Section~\ref{sec:relworks} presents related works and
Section~\ref{sec:conclusion} concludes the paper by describing
perspective work.

%% file: foss.tex
\section{\FOSS{} distributions}\label{sec:foss}

Overall, the architectures of all \FOSS{} distributions are quite
similar. Each user machine, i.e., a distribution \emph{installation},
has a local \emph{package status} recording which packages are locally
installed and which are available from remote distribution
repositories. In an \emph{upgrade scenario} the system administrator
requests a change of the package status (e.g., install, remove,
upgrade to a newer version) by the mean of a package manager, which is
in charge of finding a suitable \emph{upgrade plan}. More precisely,
the package manager solves dependencies and conflicts, retrieves
packages from remote repositories as needed, and deploys individual
packages on the filesystem, possibly aborting the operation if
problems are encountered along the way.

A \emph{package} is usually a bundle of three main parts:

\smallskip
\noindent{\bf Files} the set of files and directories shipped within
the package for installation: executable binaries, data,
documentation, etc. \emph{Configuration files} is the subset of files
affecting the runtime behavior of the package and meant to be locally
customized by the system administrator. Proper internalization of
configuration file details is relevant for our purposes, as specific
configurations can (implicitly) entail dependencies not otherwise
declared by the involved packages (see
Section~\ref{sec:elementsToBeModeled} for an example).

\smallskip
\noindent{\bf Meta-information} contains package-related
information\linebreak such as: a unique identifier, software version,
maintainer and package description, and most notably
\emph{inter-package relationships}. The kinds of relationships vary
with the distribution, but a common core subset includes: dependencies
(the need of other packages to work properly), conflicts (the
inability of being co-installed with other packages), feature
provisions (the ability to declare named features as provided by a
given package, so that other packages can depend on them), and
restricted boolean combinations of
them~\cite{edos-package-management}.

\smallskip
\noindent{\bf Maintainer scripts} are a set of programs, usually
written in shell script, that are used to enable maintainers to
attach actions to hooks that are fired by the installer. Which hooks
are available depends on the installer; \texttt{dpkg} offers one of
the most comprehensive set of hooks: pre/post-unpacking,
pre/post-removal, and upgrade/downgrade to specific
versions~\cite{debian-policy}.

Maintainer scripts are challenging objects to model, both for its
semantics (shell script is a full-fledged, Turing-complete programming
language) and for its syntax which enjoys a plethora of meta-syntactic
facilities (here-doc syntax, interpolation, etc.).

During package deployment, various kinds of failures can be induced by
maintainer scripts. The ``simplest'' example is a runtime failure of a
script (usually detected by a non-zero exit code), against which
system administrator are left helpless beside their shell script
debugging abilities. A more subtle, though possibly easier to deal
with, kind of failure are inconsistent configurations left over by
upgrade scenario not predicted by maintainers. For instance: a
maintainer script can ``forget'' to un-register a plugin from its main
application while removing the package shipping the plugin, hence
living around an inconsistent configuration (which might, or might
not, cause execution failures in the main application).


Our aim is to develop meta-models able to grasp the details of \FOSS{}
installations for the purpose of preemptive discovery of both kind of
upgrade failures. Also, in those cases where simulation is not enough
to detect failures, we want our meta-models to be able to equip
runtime execution of scripts with detailed execution logs. Those
execution logs can then be offered to state of the art roll-back
mechanisms (see Section~\ref{sec:relworks}
and~\cite{hotswup:package-upgrades}).


%% file: upgrade.tex
\begin{figure*}[t]
  \center
  \includegraphics[width=0.90\textwidth]{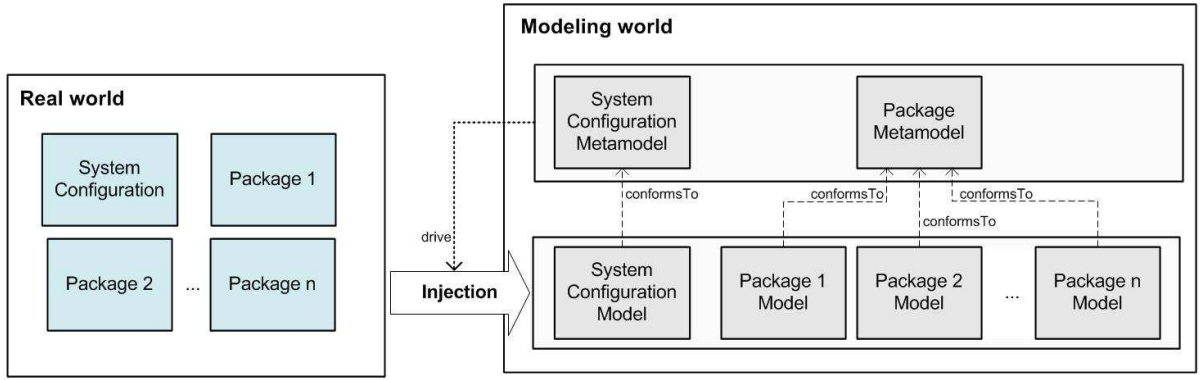}
  \caption{Model injection}
  \label{fig:injection}
\end{figure*}

\section{Model-based upgrade}\label{sec:upgrade}


The problem of maintaining \FOSS{} installations is far from trivial
and has not yet been addressed
properly~\cite{hotswup:package-upgrades}. One of the main reason is
that package managers are aware only of package meta-information
(and in particular on inter-package relationships), which are not
expressive enough. As a consequence, package managers are not enough
to detect and manage several upgrade failure scenarios.

Recently, in the context of the
\MANCOOSI\footnote{\url{http://www.mancoosi.org}} project, a
model-driven approach has been proposed~\cite{ENASE2009} to improve
upon that, by equipping package managers with model of the hosting
\FOSS{} installation. Equipped with that, package managers can both
simulate upgrades (trying to detect configuration inconsistencies)
and, during deployment on the real system, create a more detailed log
of script executions that can be used later on to pinpoint upgrade
roll-back mechanism to the precise point where the failure occurred
during deployment.

The proposed approach is then firstly based on \emph{upgrade run
  simulation}; the simulation takes into account two models (see
Figure~\ref{fig:overallApproach}): the \emph{System Configuration
  Model} and the \emph{Package Model} (see the arrow~\CIRCREF{a}). The
former describes the state of a given system in terms of installed
packages, running services, configuration files, etc. The latter
provides information about the packages involved in the upgrade in
terms of inter-package relationships. Since a trustworthy simulation
has to consider the behavior of the maintainer scripts which are
executed during the package upgrades, the package model also specifies
an abstraction of them and of their behavior.

There are two possible simulation outcomes: \emph{not valid} and
\emph{valid} (see the arrows \CIRCREF{c} and \CIRCREF{d},
respectively). In the former case it is expected that the upgrade on
the real system will fail. Thus, before proceeding with it the problem
spotted by the simulation should be fixed. In the latter
case---\emph{valid}---the upgrade on the real system can be operated
(see the arrow \CIRCREF{i}). However, since the models are an
abstraction of the reality, upgrade failures might occur due to
reasons like I/O errors or by maintainer scripts features unaccounted
for in the modeling. During package upgrades \emph{Log Models} are
produced to store all the transitions between configurations (see
arrow \CIRCREF{b}). The information contained in the system, package,
and log models (arrows \CIRCREF{e} and \CIRCREF{f}) can then be used
in case of failures (arrow \CIRCREF{l}) when the performed changes
have to be undone to bring the system back to the previous valid
configuration (arrow \CIRCREF{g}).

In order to apply on real scenario the approach depicted in Figure~\ref{fig:overallApproach},
existing systems have to be represented in terms of models. In this respect, the availability of
\emph{injectors} is crucial since they are tools that transform software artifacts into corresponding
models in an automatic way. In particular, as shown in Figure~\ref{fig:injection}, given a real
software system and a set of packages a corresponding representation in the modeling world has to be
obtained. Since it is not possible to specify in detail every single part of systems and packages,
trade-offs between model completeness and usefulness have been evaluated. In this respect, models are
specified by using modeling constructs which are formalized in specific metamodels (see
Section~\ref{sec:metamodels}) that have been conceived during a domain analysis phase (see
Section~\ref{sec:analysis}).

Over the last years, several approaches for extracting models from
software artifacts have been proposed even though the optimal solution
which can be used for any situation does not exist
yet~\cite{ICM08}. The complexity of the problem relies on the
limitation of current lexical tools which do not provide the proper
abstractions and constructs to query code and generate models with
respect to given metamodels. Some approaches like~\cite{WK06,E06}
focus on generating metamodels from grammars but they have some
drawbacks that may restrict its usefulness, such as the poor quality
of the automatically generated metamodel~\cite{ICM08}. Approaches
like~\cite{TCS} enable the automatic generation of injectors starting
from annotated metamodels with syntactic properties. However, they do
not permit reuse of existing grammars written for well-known parser
generators. Techniques like~\cite{ICM08} propose specific languages to
query software artifacts and generate models according to specified
source-to-model transformation rules.

Several projects are under development to provide tools and
methodologies for model-driven modernization and model injection. For
instance, MoDisco~\cite{MODISCO} defines an infrastructure for
supporting model-driven reverse engineering by relying on the concept
of \emph{discoverer} which is a piece of software in charge of
analyzing part of an existing system and extracting a model using the
MoDisco's infrastructure.

\subsection{\FOSS{} distributions metamodels}\label{sec:metamodels}

The metamodels which underpin the model based approach depicted in
Figure~\ref{fig:overallApproach} have been defined according to an
iterative process consisting of two main steps: \emph{a)} elicitation
of new concepts from the domain to the metamodel; and \emph{b)}
validation of the formalization of the concepts by describing part of
real systems. The metamodels which have been defined are as follows:

\begin{itemize}
\item[--] the \emph{System Configuration metamodel}, which contains
  all the modeling constructs necessary to make the \FOSS{} system
  able to perform its intended functions. In particular it specifies
  installed packages, configuration files, services, filesystem state,
  loaded modules, shared libraries, running processes, etc. The system
  configuration metamodel takes into account the possible dependency
  between the configuration of an installed package and other package
  configurations. The ability to express such fine-grained and
  installation-specific dependencies is a significant advantage
  offered by the proposed metamodels which embody domain concepts
  which are not taken into account by current package manager tools;

\item[--] the \emph{Package metamodel}, which describes the relevant
  elements making up a software package. The metamodel also gives the
  possibility to specify the maintainer script behaviors which are
  currently ignored---beside mere execution---by existing package
  managers;

\item[--] the \emph{Log metamodel}, which is based on the concept of
  transactions that represent a set of statements that change the
  system configurations. Transitions can be considered as model
  transformations~\cite{Bezivin05} which let a configuration $C_1$
  evolve into a configuration $C_2$.

\end{itemize}

Log models play a key role for both \emph{preference roll-back} and
\emph{``live'' failures}. The former takes place when a user wants to
recover a previous configuration, for whatever (even non-functional)
reason. Note that the log models provide information useful to
roll-back to any previous valid configuration, not necessary a
contiguous one. The latter happens when undetected failures are
encountered while deploying upgrades on the real system. In such a
case, the information stored in the log model are exploited to
retrieve the fallacious statements and provide hints on how to
roll-back to the configuration from which the broken transaction has
started.

\begin{figure}[t]
  \center \includegraphics[width=6cm]{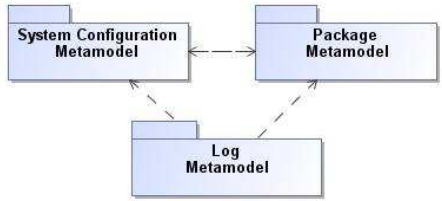}
  \caption{Dependencies among metamodels}
  \label{fig:metamodels}
\end{figure}

\noindent As shown in Figure~\ref{fig:metamodels}, \emph{System
  Configuration} and \emph{Package} metamodels have mutual
dependencies, whereas the \emph{Log} metamodel has only direct
relations with both \emph{System Configuration} and \emph{Package}
metamodels. For a detailed description of the metamodels, please refer
to~\cite{ENASE2009,mancoosi-wp2d1}.

The first step that needs to be performed when defining a metamodel is
to accurately study the domain in order to understand the elements and
the artifacts that need to be modeled as presented in the next
section.

%% file: analysis.tex
\section{Maintainer scripts analysis}\label{sec:analysis}


The metamodels outlined in Section~\ref{sec:metamodels} have been obtained though a suitable domain
analysis study. The most difficult part of this analysis process is the study of the maintainer
scripts. The adopted scripting languages are mainly POSIX shell but they are written also in
Perl~\cite{PERL}, Bash~\cite{BASH}, etc. Scripting languages have rarely been formally investigated
and with no exciting results~\cite{aiken-sql-injection-detection,mazurak-abash}, thus posing
additional difficulties in understanding their side-effects which can spread throughout the whole
(file)system. Our aim is to describe the most common macro-actions of maintainer scripts in terms of
models which abstract from the real system, but are expressive enough to grasp several of their
effects on package upgrades.
The analysis phase is then extremely important in order to find the right trade-off between
expressiveness and abstraction.

Due to the large amount of scripts to consider (e.g., about 25$^.$000
on Debian Lenny, see Section~\ref{sec:scriptAnalysis}), we tried to
collect them in clusters to be able to concentrate the analysis on
representatives of the equivalence classes identified. The adopted
procedure for clustering has been as follows:

\begin{enumerate}

\item \emph{Collect all maintainer scripts} of a given
  distribution. By not choosing a random subset we are sure to have
  collected the most representative set of scripts;

\item \emph{Identify scripts generated from helper tools}. A large
  number of scripts or part of them is generated by means of
  ``helper'' tools that provide a collection of small, simple and
  easily understood tools that are used to automate various common
  aspects of building a package. Since (part of) maintainer scripts
  are automatically generated using these helpers and their boiler
  plates, we can concentrate the analysis on the helpers themselves,
  rather than on the result of their usage;

\item \emph{Ignore inert script parts}. As all scripting languages,
  shell scripts contain parts which do not affect their computational state such as blank lines or comments.  Intertwined with the
  removal of generated parts (to be analyzed later on) we have
  systematically ignored inert script parts, possibly leading upon
  removal to empty scripts that have been therefore ignored as a
  whole;

\item \emph{Study of scripts written ``by hand''}. The remaining
  scripts need to be more carefully studied, as they have been written
  from scratch by package maintainers to address a specific need, most
  likely not covered by any helper tool. Actually we worked on
  identifying further recurrent templates that maintainers use when writing
  the scripts.

\end{enumerate}

In the remainder of this section we present the result of the analysis performed on two
representative FOSS distributions: Debian GNU/Linux and an RPM-based distribution. Note that due to
space restrictions we cannot report all the details, but the interested reader can refer
to~\cite{mancoosi-wp2d1}.

\subsection{Debian GNU/Linux}
\label{sec:scriptAnalysis}

The analysis has been performed considering a ``snapshot'' of Debian
\emph{Lenny}, the just released ``stable'' brand of Debian. The
snapshot has been taken on December 4th, 2008, considering only the
\texttt{amd64} architecture (soon to become the most widespread
architecture on end-user machines), and all the packages shipped by
the Debian archive and targeted at the end user (i.e., sections
\texttt{main}, \texttt{contrib}, and \texttt{non-free}). Each (binary)
package\footnote{From now on, unless otherwise stated, we will use the
  term ``package'' to refer to binary packages.}  in Debian can come
with 5 different kinds of maintainer scripts:

\begin{table*}[htbp]
  \begin{center}
    {
      \begin{tabular}{|c|c|l|}
        \hline
        \textbf{Group} & \textbf{Occurrences} & \textbf{Representative script name} \\
        \hline
        G1 & 93 & libk/libkpathsea4\_2007.dfsg.2-4\_amd64.deb.preinst \\
        \hline
        G2 & 54 & d/dict-freedict-swe-eng\_1.3-4\_all.deb.postinst \\
        \hline
        G3 & 54 & d/dict-freedict-fra-deu\_1.3-4\_all.deb.postrm\\
        \hline G4 & 35 & j/jabber-jud\_0.5-3+b1\_amd64.deb.preinst\\
        \hline G5 & 35 & g/gauche-c-wrapper\_0.5.4-2\_amd64.deb.postinst\\
        \hline G6 & 33 & w/wogerman\_2-25\_all.deb.config\\
        \hline G7 & 31 & m/mii-diag\_2.11-2\_amd64.deb.prerm\\
        \hline G8 & 30 & libs/libsocket6-perl\_0.20-1\_amd64.deb.postrm\\
        \hline
        $\cdots$ & $\cdots$ & \hspace{0.5cm}$\cdots$
    \end{tabular}}
    \caption{Excerpt of the obtained groups} \label{tab:groups}
  \end{center}
\end{table*}

\begin{enumerate}
\item \PREINST{} (mnemonic for ``pre-installation'') scripts that are
  run before the files shipped by a package being installed have been
  unpacked on the filesystem of the target machine;


\item \POSTINST{} (mnemonic for ``post-installation'') scripts that are
  run after the files shipped by a package have been unpacked on the
  target filesystem;


\item \PRERM{} (mnemonic for ``pre-removal script'') scripts that are
  executed just before removing from the target filesystem those files
  which belong to the package which is being removed;


\item \POSTRM{} (mnemonic for ``post-removal'') scripts that are
  executed just after removing the files belonging to the package
  being removed from the filesystem;


\item \CONFIG{} (mnemonic for ``configuration'') scripts that are used
  to configure a software which requires specific user input to be
  configured.


\end{enumerate}

Considering 5 maintainer scripts per package we obtain a potential universe of scripts to be
considered of 114$^.$115 scripts (i.e., 22$^.$823 $\times$ 5). Luckily, 88$^.$675 (77.7\%) of those
are actually missing from the corresponding packages. This means that the remaining actual script
universe which need to be analyzed consists of ``just'' 25$^.$440 scripts (22.3\% of 114$^.$115).

\subsubsection{Scripts generated from helpers}

Package maintainers use complex toolchains to facilitate their maintenance work which is otherwise
prone to repetition of self-similar tasks. In Debian, the legacy helpers used to generate maintainer
script snippets is the \texttt{debhelper} collection~\cite{debhelper}. For the most part, it consists
of tools which are invoked at package build time to automate package-construction tasks such as
installing specific file categories (e.g., manual pages, documentation, etc.) in the location
prescribed by the Debian policy~\cite{debian-policy}. Instead of requiring each maintainer to write
exactly the same shell script snippets by hand, \texttt{debhelper} also offers a template mechanism
called ``autoscripts'' which writes down the needed snippets when needed.
To our ends, this means that we can restrict our analysis to the
templates themselves, because they are either verbatim copied in the
resulting scripts or, in the worst case scenario, filled using simple
textual ``holes'' such as the current package names.

We extracted from \texttt{debhelper} autoscripts templates, which amount to 52 templates in
total~\cite{mancoosi-wp2d1}. Each of those templates contains statements that are executed as a
whole. For instance, in Listing~\ref{lst:postrm-gconf} a sample template is reported. It consists of
statements which are executed after the removal of the GNOME configuration tool~\footnote{GNOME: The
Free Software Desktop Project - \url{http://http:www.gnome.org}}.


\begin{lstlisting}[breaklines,style=AMMA,language=SH,mathescape=false,rulesepcolor=\color{black},caption={Sample  tamplate}, label={lst:postrm-gconf}]
if [ "$1" = purge ]; then
    OLD_DIR=/etc/gconf/schemas
    SCHEMA_FILES="#SCHEMAS#"
    if [ -d $OLD_DIR ]; then
        for SCHEMA in $SCHEMA_FILES; do
            rm -f $OLD_DIR/$SCHEMA
        done
        rmdir -p --ignore-fail-on-non-empty $OLD_DIR
    fi
fi
\end{lstlisting}

All the recurrent templates like the one in Listing~\ref{lst:postrm-gconf} are formalized in the
metamodel~\cite{mancoosi-wp2d1}. In this way maintainers, which are used to write scripts by means of
\texttt{debhelper} and the previously identified templates, will find familiar and meaningful
statements in the metamodel.

When the maintainer needs to add specific code to maintainer scripts,
code which is not provided by autoscript templates, \texttt{debhelper}
enables mixing generated lines with lines written by hand. All
generated lines are tagged with specially-crafted comments, so that
they are recognizable mechanically.

Starting from the non-empty maintainer scripts extracted from Lenny
(summing up to 25$^.$440 scripts), we analyzed how many of them are
\emph{entirely} composed by lines generated using the autoscript
mechanism. Also, we produced a ``filtered'' version of all the
remaining maintainer scripts (i.e., those that contain at least
\emph{some} line written by hand by the package maintainer) which has
been analyzed later on in more details. The summary of generated (part
of) maintainer scripts is as follows:

{
\begin{center}
  \begin{tabular}{l|c|c}
    & \multicolumn{1}{c|}{\emph{n. of scripts}}
    & \multicolumn{1}{c}{\emph{lines of code (LOCs)}} \\\hline
    non-blank  & 25$^.$440 (100\%)  & 386$^.$688 (100\%)  \\
    generated  & 16$^.$348 (64.3\%) & 162$^.$074 (41.9\%) \\
    by hand    & 9$^.$061 (35.6\%)  & 224$^.$614 (58.1\%) \\
  \end{tabular}
\end{center}}

About $2/3$ of all the maintainer scripts are composed only of lines
generated using the autoscript mechanism.

\subsubsection{Analysis of scripts ``by hand''}

The scripts that survived to the previous phases are 9$^.$061. These
scripts have been analyzed ``by hand''.  The idea of that final
analysis is to find additional templates or additional statements that
should be considered when defining the metamodel.  The analysis ``by
hand'' has been performed as follows:

\begin{table}[htbp]
  \begin{center}
    {
      \begin{tabular}{|c|c|c|}
        \hline \textbf{Template} & \textbf{Occurrences} & \textbf{Origin Group}\\
        \hline
        Template1 & 97 & G1\\
        \hline
        Template2 & 69  & G14\\
        \hline $\cdots$ & $\cdots$ & $\cdots$
    \end{tabular}}
    \caption{Excerpt of the occurrences of the Templates}
    \label{tab:templates}
  \end{center}
\end{table}

\begin{enumerate}

\item All the scripts that survived to the previous pruning phases are
  clustered in groups, where a group collects scripts that contain
  exactly the same statements. For each group we then selected one
  representative.  Table~\ref{tab:groups} shows an excerpt of the
  groups that we identified, ordered by occurrence. The second column
  shows the occurrences while the third column contains the name of the
  script representative of the group. By referring to
  Table~\ref{tab:groups}, group G1 consists of 92 scripts which are identical to the
  script
  \texttt{libk/libkpathsea4\_2007.}\texttt{dfsg.2-4\_amd64.deb.preinst}
  and then this script identifies a potential template;

\item The next step consists in identifying the occurrence of the
  identified templates in the collection of 9$^.$061 scripts. For
  instance the occurrence of \emph{Template1} is 97 and refers to
  group \emph{G1} of Table~\ref{tab:groups}. Table~\ref{tab:templates}
  shows an excerpt of the template occurrences. More precisely, the
  first column contains the elicited templates and the second one their occurrences. Last column
  refers to the groups which originated the considered templates;

\item Once templates have been identified (we identified 116 templates)
  together with their occurrences, the next step consists
  of identifying similarities among templates in order to collect them
  in classes. In fact, we recall that the occurrences are calculated
  with exact matching and that a white space can also compromise the
  matching. We found 10 classes that collect 1$^.$340 scripts;

\item The next step consists in analyzing each class in order to
  understand how to deal with this kind of scripts. In other words, we
  have to understand whether the already identified statements are
  sufficient or whether new statements are required;

\item The last step has been the analysis of scripts ``by hand'' with
  occurrence 1 in order to understand whether they are already covered
  or whether they contain statements that we are not able to deal
  with. In this last step we analyzed other 43 scripts.

\end{enumerate}

Summarizing the entire process, the total amount of considered packages is 22$^.$823. Considering 5
maintainer scripts per package we obtain a (potential) universe of 114$^.$115 scripts (i.e.,
22$^.$823 $\times$ 5). Luckily, 88$^.$675 (77.7\%) of those are actually missing from the
corresponding packages.  The universe of the remaining scripts consists of 25$^.$440 scripts
(22.3\%). The analysis covered approximatively the 66\% of the existing 25$^.$440 scripts and the
93\% of the universe of 114$^.$115 potential scripts.

\begin{figure*}[t]
  \center
  \begin{tabular}{c c}
    \includegraphics[width=9cm]{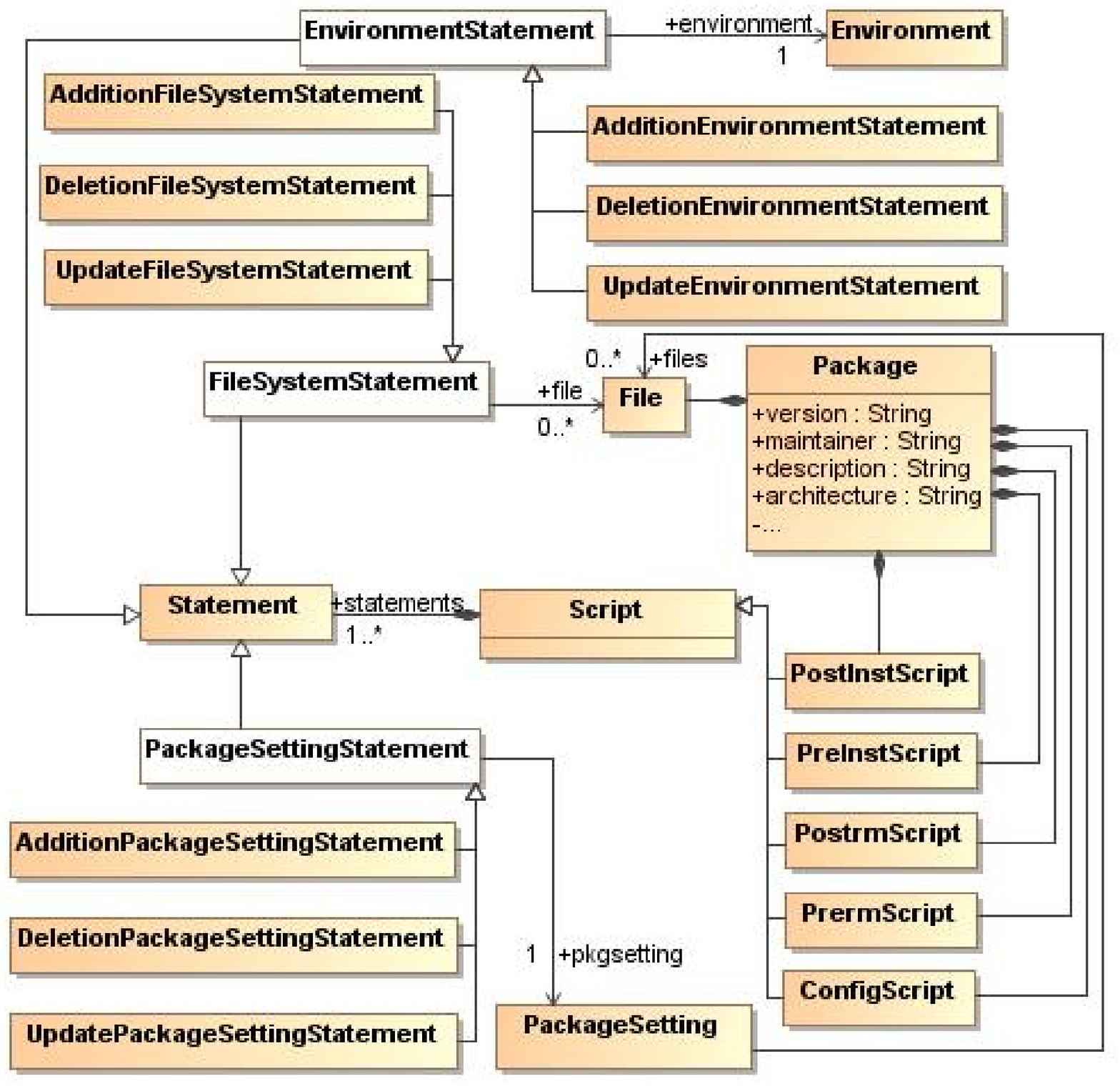}
     & \raisebox{1cm}{\includegraphics[width=5cm]{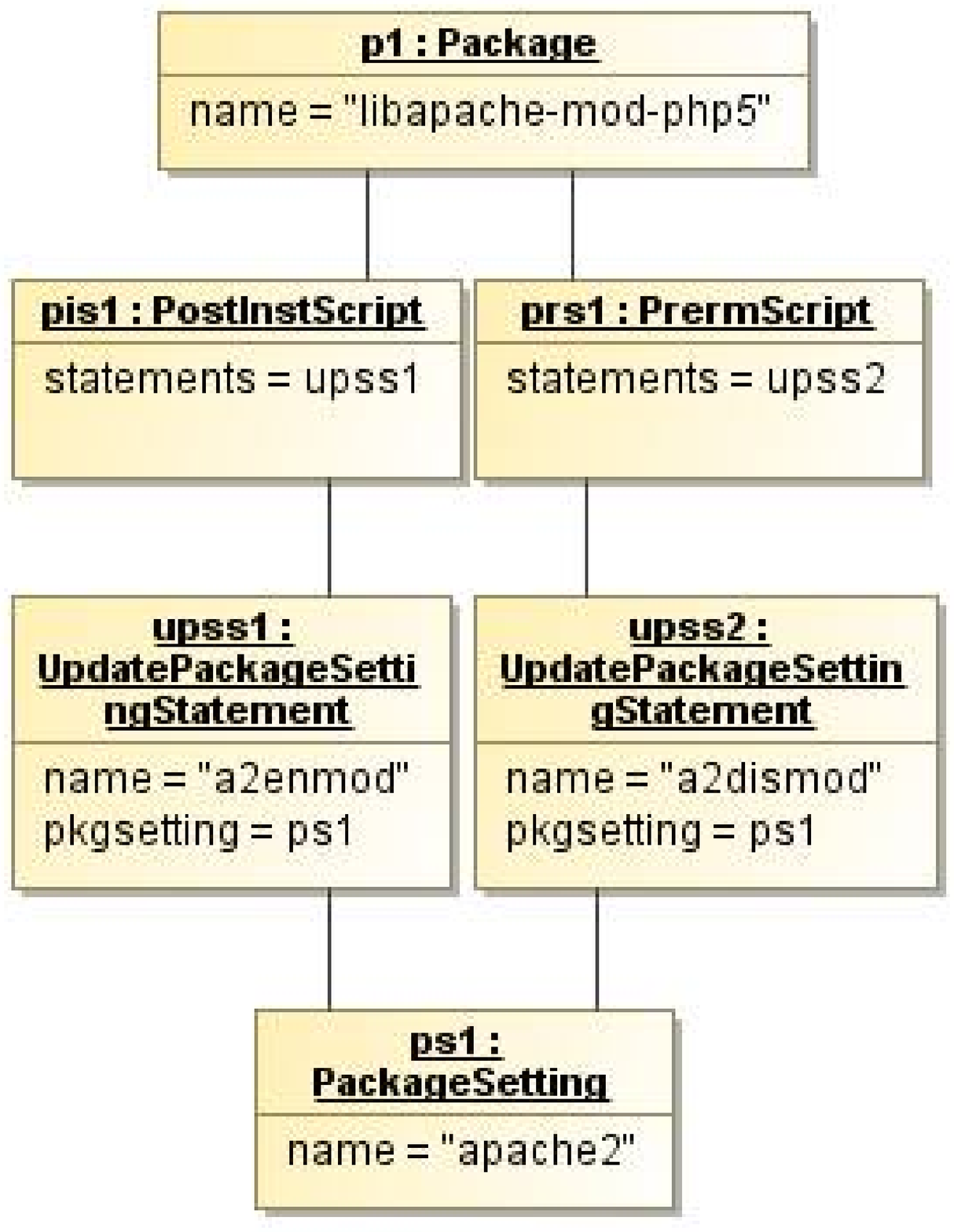}}
    \\
    a) Fragment of the Package metamodel
     & b) Sample Package model \\
  \end{tabular}
  \caption{Package specification}
  \label{fig:packageMetamodel}
\end{figure*}

\subsection{RPM-based distributions}\label{sec:analysisRPM}

RPM (RPM Package Manager) is one of the most common software package manager used within \FOSS{}
distributions. Although RPM was originally designed to work with Red Hat Linux, it is nowadays used
in several other distributions, such as Mandriva, Fedora, and Suse.

The \SPEC~\cite{RPMBook} file plays the main role in the RPM package build process. In fact, such
file contains all the information required to (i) compile the program and build source and binary
packages, and (ii) install and uninstall the program on the target machine. The \SPEC{} file is
divided in several sections and each section is denoted by a corresponding keyword like
\texttt{\%pre}.

Since we are interested in installation and removal aspects, we focus only on the RPM sections that
are involved: \emph{Install} and \emph{Uninstall} scripts section. Similarly to Debian, in the RPM
format there are four kind of scripts (\texttt{\%pre},~\texttt{\%post},~\texttt{\%preun}, and
~\texttt{\%postun} ) each of them meant to be executed at different stages of the package upgrade
process.

It is not very common having RPM packages that require actions to be performed prior to installation.
In fact none of the 350 packages that comprise Red Hat Linux 4.0 makes use of  \texttt{\%pre}
scripts. A typical example of \texttt{\%post} script (which is executed after installation) consists
of the \texttt{ldconfig} command which updates the list of available shared libraries after a new one
has been installed. If a package uses a \texttt{\%post} script to perform some function, quite often
it will include a \texttt{\%postun} script that performs the inverse of the \texttt{\%post} script,
after the package has been removed.

The scripts which are executed before removing packages (\texttt{\%preun}) are  used to prepare the
system immediately prior the package deletion. Specularly, \texttt{\%postun} is executed after
package deletions. Quite often, \texttt{\%postun} scripts are used to run \texttt{ldconfig} to remove
newly erased shared libraries from \texttt{ld.so.cache}. As highlighted before, these scripts
typically do the inverse of \texttt{\%post} ones (which might happen to be the same action, as with
\texttt{ldconfig}, when it simply consists in updating some sort of cache/registry).

\smallskip
 Similarly to Debian, Fedora,\footnote{Fedora Project Web site:
  \url{http://fedoraproject.org}} an RPM-based Linux distribution,
also makes use of templates for the maintainer scripts\footnote{Fedora
  scriptlet snippets: \url{http://fedoraproject.org/}
  \url{wiki/Packaging/ScriptletSnippets}}. Such templates, called
autoscripts, are reported in~\cite{mancoosi-wp2d1}. The \SPEC{} files of the Fedora distribution we
have considered can be downloaded at \url{http://svn.rpmforge.net/} \url{svn/trunk/rpms/}. The
available \SPEC{} files are 4$^.$704, and considering that each of them can contain four kinds of
scripts, the potential universe that has to be analyzed consists of 4$^.$704*4=18$^.$816 scripts.
Actually, the scripts that are present in this set are 2$^.$038, that is approximately 10.8\% of
18$^.$816. These scripts are divided as follows: 81 \%\texttt{pre}, 911 \%\texttt{post}, 234
\%\texttt{preun}, and 812 \%\texttt{postun}. We extracted the four kinds of scripts from each spec
file and, in order to make the analysis, we created four new files containing the scripts.

Unfortunately, in this case we are not able to identify scripts that
are generated from helpers, since there is no marking that helps in
identifying the generated parts. For this reason, we performed
analysis ``by hand'', similarly to what is described for Debian. Then,
all the scripts are clustered in groups, where a group collects
scripts that contain exactly the same statements. For each group we
then selected one representative.

From this first analysis we can say that more than 50\% of the scripts
can be generated by templates.  Once the scripts previously identified
have been deleted, we continued the analysis, the next step being to
use the identified templates in order to check their matches as part
of the code of a script. Since the matches are always exact, we
performed another step of analysis. We then manually inspected the
scripts and manually identified the match.  Thus, we defined other
templates that complement the already defined templates.

To summarize, this analysis demonstrates that, by means of templates,
1$^.$962 scripts among the 2$^.$038 that constitute our universe of
scripts can be automatically generated for sure ($\sim$93,6\%). Please
also remember that the ``potential'' amount of scripts, as described
at the beginning of this section, is the number of spec files
multiplied by 4 (that is the number of different kinds of
scripts). Then the total amount of potential scripts is 18$^.$816, and
the remaining scripts, which are 76, represent the 0,4\% of the total
potential. Furthermore, a large part of the remaining 76 scripts is
simply a combination of some customized templates or parts of
templates that could be modeled by means of more simple statements.

\begin{figure*}[thb!]
  \center
  \includegraphics[width=17.5cm]{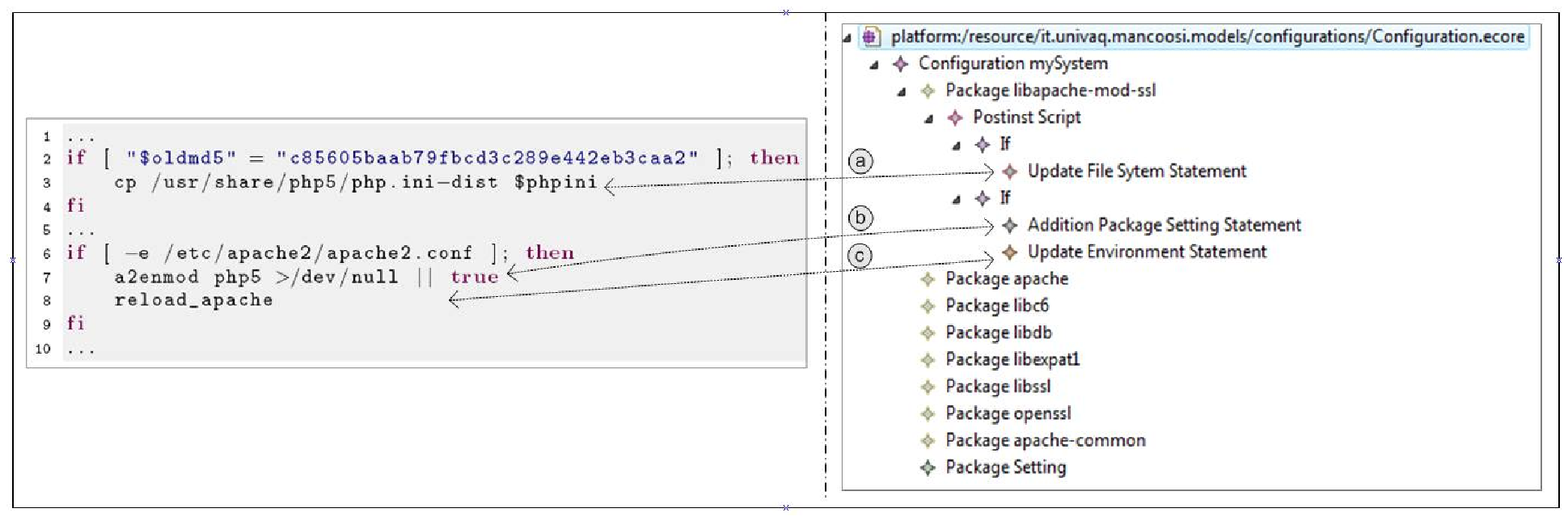}
  \caption{Fragment of the
    \texttt{libapache2-mod-php5\_5.2.6-5\_amd64.deb.postinst} script}
  \label{fig:scriptCodeModel}
\end{figure*}

%% file: metamodel.tex
\section{Modeling maintainer scripts}\label{sec:elementsToBeModeled}

\noindent The outcome of the analysis has been concretized in the metamodels summarized in
Sect.~\ref{sec:metamodels}. They have been validated with real systems and in the following we report
an example which relies on the \texttt{Package} metamodel depicted in
Figure~\ref{fig:packageMetamodel}.a. It contains metaclasses that are required to model maintainer
scripts like the following:

\begin{lstlisting}[breaklines,style=AMMA,language=SH,mathescape,rulesepcolor=\color{black},label={lst:postinst}]
#!/bin/sh
if [ -e /etc/apache2/apache2.conf ] ; then
    a2enmod php5 >/dev/null || true
    reload_apache
fi
\end{lstlisting}

In particular, the installation of PHP5---a web scripting language
integrated with the Apache web server---executes the \emph{postinst}
script above which will be executed once the package has been unpacked
on the target system. Essentially, the installed Apache module
\texttt{php5} gets enabled by the above snippet invoking the
\texttt{a2enmod} command in line 3. The Apache service is then
reloaded (line 4) to make the change effective. Upon PHP5 removal, the
reverse will happen, as implemented by the following \PRERM{} script
pertaining to the PHP5 package:

\begin{lstlisting}[breaklines,style=AMMA,language=SH,mathescape,rulesepcolor=\color{black},label={lst:prerm}]
#!/bin/sh
if [ -e /etc/apache2/apache2.conf ] ; then
    a2dismod php5 || true
fi
\end{lstlisting}

The model-based specification of such scripts encompasses the metaclass \texttt{Statement} in
Figure~\ref{fig:packageMetamodel}.a which represents an abstraction of the commands that can be
executed by a given script to affect the environment, the file system or the package settings of a
given configuration (\texttt{EnvironmentStatement}, \texttt{FileSystemStatement}, and
\texttt{PackageSettingStatement}, respectively). For instance, the sample model in
Figure~\ref{fig:packageMetamodel}.b reports the scripts contained in the package
\texttt{libapache-mod-php5} which contains the scripts reported at the beginning of the present
section. For clarity of presentation, Figure~\ref{fig:packageMetamodel}.b contains only the relevant
elements of the \POSTINST{} and \PRERM{} scripts which are represented by the elements \texttt{pis1}
and \texttt{prs1}, respectively.

%
%

The most significant metaclasses of the Package metamodel which underpin the script behavior
specification are \texttt{Script}, and \texttt{Statement}.
%
%
In particular, according to the different configuration elements which can be affected by the
execution of a given script statement, the abstract metaclass \texttt{Statement} is specialized in
different metaclasses that are \texttt{FileSystemStatement}, \texttt{EnvironmentStatement}, and
\texttt{Package\-Setting\-Statement}. Moreover, each of them is in turn specialized for capturing
additions, removals, and upgrades. In particular, the statements which add, delete, and modify the
\texttt{FileSystem} are respectively represented as \texttt{Addition\-FileSystem\-Statement},
\texttt{DeletionFileSystem\-Statement} and \texttt{Update\-File\-System\-Statement} instances. The
shell commands \texttt{touch}, \texttt{rm} and \texttt{cp}, are sample instances of such metaclass.

The statements which modify the \texttt{Environment} of a given configuration are given in terms of
instances of \texttt{EnvironmentStatement} specializations. Shell commands like
\texttt{install-menu}, \texttt{rmmod}, \texttt{ldconfig} of Linux distributions, can be respectively
modeled as \texttt{Addition\-Environment\-Statement}, \texttt{Deletion\-Environment\-Statement} and
\texttt{Update\-Environment\-Statement} instances.

As pointed out in Section~\ref{sec:foss}, an installed package might depend on settings properly
stored in dedicated configuration files (i.e., the service \texttt{apache2} depends on the
configurations specified in the file \texttt{httpd.conf} usually stored in the \texttt{/etc/apache2}
directory). The statements which modify such settings are modeled by means of instances of the
\texttt{PackageSettingStatement} extensions. Finally, maintainer scripts might contain statements
which do not change the system configuration but are comments, emit messages, etc. Such cases can be
specified by means of instances of the \texttt{NeutralStatement} metaclass.

A summarizing example is depicted in Figure~\ref{fig:scriptCodeModel} which reports a fragment of the
\texttt{postinst} script of the Debian Lenny \texttt{libapache2-mod-php5\_5.2.6-5\_amd64.deb}
package. The code is injected with respect to the \texttt{Package} metamodel summarized above by
giving place to the model on  the right-hand side of Figure~\ref{fig:scriptCodeModel}. In particular,
the copy operation of the file \texttt{php.ini-dist} represents a modification of the file system and
is hence\index{} modeled as an \texttt{UpdateFileSystemStatement} element (see the
arrow~\CIRCREF{a}). Once the \texttt{php5} module has been installed, the configuration of the
\texttt{apache2} package has to be modified by enabling the new module. This operation is performed
by executing the command \texttt{a2enmod} which is modeled as an
\texttt{AdditionPackageSettingStatement} element (see the arrow \CIRCREF{b}). Finally, the
\texttt{UpdateEnvironmentStatement} element in the model (see the arrow \CIRCREF{c}) represents the
command which reloads the \texttt{Apache} Web server to update the environment with the previous
modification.

%% file: related.tex
\section{Related works}\label{sec:relworks}

The main difficulties related to the management of upgrades in \FOSS{} distributions depend on the
existence of maintainer scripts which can have system-wide side-effects, and hence can not be
narrowed to the involved packages only. An interesting proposal to support the upgrade of a system,
called NixOS, is presented in~\cite{dolstra-nixos}. NixOS is a purely functional distribution meaning
that all static parts of a system (such as software packages, configuration files and system boot
scripts) are expressed as pure functions. Among the main limitations of NixOS there is the fact that
some actions related to upgrade deployment can not be made purely functional (e.g., user database
management). \cite{McQ2005} proposes an attempt to monitor the upgrade process with the aim to
discover what is actually being touched by an upgrade. Unfortunately, it is not sufficient to know
which files have been involved in the maintainer scripts execution but we have also to consider
system configuration, running services etc., as taken into account by our metamodels.

Concerning software modernization, OMG defined the Architecture Driven Modernization (ADM) task
force~\cite{ADM} that aims at building standard metamodels and tools for supporting software renewal.
Reus et al. in~\cite{RGD06} and~\cite{fleurey07a} propose similar MDA processes for software
migration. They parse the text of the original system and build a model of the abstract syntax tree.
This model is then transformed into a pivot language that can be translated into UML. The context of
such works is different from that considered in this paper which deals with \FOSS{} distributions and
especially with maintainer scripts.

%% file: conclusions.tex
\section{Conclusion and future work}\label{sec:conclusion}

Dealing with upgrade failures in \FOSS{} systems is a challenging
task, mainly due to the complexity of maintainer scripts which are
executed during upgrade deployment on the real system. Such scripts
are written in languages that have rarely been formally investigated,
thus posing additional difficulties in understanding their system-wide
side-effects.

In this paper we have analyzed two \FOSS{} distributions and outlined
the corresponding metamodels which have been conceived to support a
model-driven approach for simulating upgrades and equipping roll-back
mechanisms with detailed deployment logs, including maintainer script
actions. The benefits of such an approach are manifold:
\begin{enumerate}
\item consistency checking possibilities are increased and trustworthy
  simulations of package upgrades become easier with respect to
  current package managers which only take into account inter-package
  relationships;
\item models can drive roll-back operations to recover previous
  configurations according to user decisions or forced by upgrade
  failures;
\item the evolution and the modifications the system underwent during
  its life cycle can consistently be recorded and used at run-time for
  roll-back operations.
\end{enumerate}
This way installation and removal simulations can take into account both package dependencies and the
behavior of maintainer scripts (which currently, on real systems, are used at deployment-time but
ignored for planning) leading to more realistic simulations and enabling checking for more complex
model inconsistencies. Even tough in this paper only two Linux distributions have been taken into
account, the proposed approach is meant to be general. In this respect, the proposed metamodels will
be refined by means of an iterative approach in order to capture unforeseen elements which are
required to specify FOSS systems in general. The evolution of metamodels gives place to a number of
problems related to the management of existing models which have to be adapted once the corresponding
metamodels change. In order to deal with such problems, the approach in~\cite{CDEP08-2,CDP08} will be
taken into account and used in the domain of FOSS systems.

As claimed in different parts of this work, the extraction of models from legacy software artifacts
is a challenging task and existing lexical tools can not be directly used because of their limited
capabilities to ``query'' textual artifacts and generate corresponding models. This represents the
most important future work which has to be carried on to fully support maintainer script
modernization. Existing approaches (like~\cite{MODISCO} and~\cite{TCS}) will be considered and
applied to the metamodels which have been conceived according to the analysis presented in this work.
Once the injection phase has been sufficiently automatized, we will instantiate the metamodels on a
widely used \FOSS{} distribution, and develop a supporting tool for integrating the presented
model-driven approach with the existing system configuration/management tools. Moreover, we will
define a model-based language, to be substituted to existing ones, for specifying the system
configurations and packages at a higher level of abstraction. This new language will enable
simulation and verification of maintainer scripts and will drive the roll-back of system upgrades.

On the practical side, we currently have the metamodels, and in
particular the log metamodel, but we still lack the communication
infrastructure with a package manager on one side and a log-equipped
roll-back mechanism. We are closing this gap together with other
partners of the \MANCOOSI{} project. The aim is to shortly have a
prototype of a next-generation package manager which couples detailed
knowledge of the installation via models with roll-back capabilities
driven by upgrade deployment logs.

%% file: iwoce2009.bbl
\begin{thebibliography}{10}

\bibitem{apache-maven}
{Apache Software Foundation}.
\newblock {Maven} project.
\newblock \url{http://maven.apache.org/}, 2009.

\bibitem{RPMBook}
{Edward C.} Bailey.
\newblock {\em Maximum {RPM}: Taking the Red Hat Package Manager to the Limit}.
\newblock {Red Hat} software, 1997.

\bibitem{Bezivin05}
J.~B\'ezivin.
\newblock {On the Unification Power of Models}.
\newblock {\em SOSYM}, 4(2):171--188, 2005.

\bibitem{CDEP08-2}
A.~Cicchetti, D.~{Di Ruscio}, R.~Eramo, and A.~Pierantonio.
\newblock Automating co-evolution in model-driven engineering.
\newblock In {\em 12th IEEE International EDOC Conference (EDOC 2008)}, pages
  222--231, M{\"u}nchen (Germany), 2008. IEEE Computer Society.

\bibitem{ENASE2009}
Antonio Cicchetti, Davide {Di Ruscio}, Patrizio Pelliccione, Alfonso
  Pierantonio, and Stefano Zacchiroli.
\newblock Towards a model driven approach to upgrade complex software systems.
\newblock In {\em Proceedings of ENASE}, 2009.

\bibitem{CDP08}
Antonio Cicchetti, Davide~Di Ruscio, and Alfonso Pierantonio.
\newblock Managing model conflicts in distributed development.
\newblock In {\em MoDELS 2008}, volume 5301 of {\em LNCS}, pages 311--325,
  2008.

\bibitem{debhelper}
Debian.
\newblock \texttt{debhelper} package.
\newblock \url{http://packages.debian.org/lenny/debhelper}, 2009.

\bibitem{hotswup:package-upgrades}
Roberto {Di Cosmo}, Paulo Trezentos, and Stefano Zacchiroli.
\newblock Package upgrades in {FOSS} distributions: details and challenges.
\newblock In {\em HotSWUp'08}, pages 1--5. ACM, 2008.

\bibitem{mancoosi-wp2d1}
Davide {Di Ruscio}, Patrizio Pelliccione, Alfonso Pierantonio, and Stefano
  Zacchiroli.
\newblock Metamodel for describing system structure and state.
\newblock {Mancoosi Project} deliverable D2.1, January 2009.
\newblock \url{http://www.mancoosi.org/deliverables/d2.1.pdf}.

\bibitem{dolstra-nixos}
Eelco Dolstra and Andres L\"oh.
\newblock {NixOS}: A purely functional linux distribution.
\newblock In {\em ICFP}, 2008.
\newblock To appear.

\bibitem{MODISCO}
Eclipse.
\newblock Modisco project.
\newblock Available: http://www.eclipse.org/gmt/modisco/.

\bibitem{edos-package-management}
{EDOS Project}.
\newblock Report on formal management of software dependencies.
\newblock EDOS Project Deliverable D2.1 and D2.2, March 2006.

\bibitem{E06}
S.~Efftinge.
\newblock openarchitectureware 4.1 xtext language reference, August 2006.
\newblock http://www.eclipse.org/gmt/oaw/doc/4.1/r80 xtextReference.pdf.

\bibitem{fleurey07a}
Franck Fleurey, Erwan Breton, Benoit Baudry, and Alain Nicolasand Jean-Marc
  Jézéquel.
\newblock Model-driven engineering for software migration in a large industrial
  context.
\newblock In {\em MoDELS'07}, 2007.

\bibitem{debian-policy}
Ian Jackson and Christian Schwarz.
\newblock Debian policy manual.
\newblock \url{http://www.debian.org/doc/debian-policy/}, 2008.

\bibitem{ICM08}
{J.L.C. Izquierdo}, {J.S. Cuadrado}, and {J.G. Molina}.
\newblock {Gra2MoL}: A domain specific transformation language for bridging
  grammarware to modelware in software modernization.
\newblock In {\em Workshop on Model-Driven Software Evolution}, 2008.

\bibitem{TCS}
Frédéric Jouault, Jean Bézivin, and Ivan Kurtev.
\newblock {TCS}: a {DSL} for the specification of textual concrete syntaxes in
  model engineering.
\newblock In {\em Proceedings of GPCE'06}, pages 249--254. ACM, 2006.

\bibitem{mazurak-abash}
Karl Mazurak and Steve Zdancewic.
\newblock Abash: finding bugs in bash scripts.
\newblock In {\em PLAS '07}, pages 105--114. ACM, 2007.

\bibitem{McQ2005}
Robert McQueen.
\newblock Creating, reverting \& manipulating filesystem changesets on {Linux}.
\newblock May 2005.

\bibitem{apt-howto}
Gustavo {Noronha Silva}.
\newblock {APT} howto.
\newblock \url{http://www.debian.org/doc/manuals/apt-howto/}, 2008.

\bibitem{ADM}
{Object Management Group (OMG)}.
\newblock {\em Architecture-Driven Modernization (ADM)}.
\newblock http://adm.omg.org/.

\bibitem{RGD06}
{T. Reus}, {H. Geers}, and {A. van Deursen}.
\newblock Harvesting software systems for mda-based reengineering.
\newblock In {A. Rensink} and {J. Warmer}, editors, {\em ECMDA-FA 2006}, volume
  4066, page 213–225. Springer LNCS, 2006.

\bibitem{BASH}
{The Free Software Foundation}.
\newblock Bash shell.
\newblock \url{http://www.gnu.org/software/bash/}, 2009.

\bibitem{PERL}
{The Perl Foundation}.
\newblock The perl directory.
\newblock \url{http://www.perl.org/}, 2009.

\bibitem{WK06}
M.~Wimmer and G.~Kramler.
\newblock {Bridging grammarware and modelware}.
\newblock In {\em Satellite Events at the MoDELS 2005 Conference}, volume 3844
  of LNCS, pages 159--168. Springer-Verlag, 2006.

\bibitem{aiken-sql-injection-detection}
Yichen Xie and Alex Aiken.
\newblock Static detection of security vulnerabilities in scripting languages.
\newblock In {\em USENIX-SS'06}, pages 179--192, 2006.

\end{thebibliography}
